# Noble gas as a functional dopant in ZnO

Oleksandr I. Malyi[1*], Kostiantyn V. Sopiha[2], and Clas Persson[1]

1 – Centre for Materials Science and Nanotechnology, Department of Physics, University of Oslo, P.O. Box 1048 Blindern, NO-0316 Oslo, Norway

2 – Ångström Solar Center, Solid State Electronics, Department of Engineering Sciences, Uppsala University, Box 534, SE-75121 Uppsala, Sweden

email: oleksandrmalyi@gmail.com (O.I.M)

**ABSTRACT**

Owing to fully occupied orbitals, noble gases are considered to be chemically inert and to have limited effect on materials properties under standard conditions. However, using first-principles calculations, we demonstrate herein that the insertion of noble gas (*i.e.*, He, Ne, or Ar) in ZnO results in local destabilization of electron density of the material driven by minimization of an unfavorable overlap of atomic orbitals of the noble gas and its surrounding atoms. Specifically, the noble gas defect (interstitial or substitutional) in ZnO pushes the electron density of its surrounding atoms away from the defect. Simultaneously, the host material confines the electron density of the noble gas. As a consequence, the interaction of He, Ne, or Ar with O vacancies of ZnO in different charge states $q$ (ZnO:$V_O^q$) affects the vacancy stability and their electronic structures. Remarkably, we find that the noble gas is a functional dopant that can delocalize the deep in-gap $V_O^q$ states and lift electrons associated with the vacancy to the conduction band.

## INTRODUCTION

Understanding the role of point defects in materials is the paramount issue in modern solid-state physics/chemistry.[1, 2] With the rapid development of experimental and theoretical methods, one already has some knowledge on defects in many technologically important binary metal oxides and can often foresee stabilization of intrinsic donor and acceptor defects under different oxygen partial pressures. However, it appears that defect characterization and especially assigning experimentally determined materials properties to a specific defect are a formidable challenge.[3] The main problem arises from a large variety of defects which can be formed. Moreover, many materials can exhibit self-doping caused by low formation energies of the defects involving impurities that exist under typical growth or characterization conditions. Specifically, it has been shown that n- or p-type conductivity of many metal oxides can be caused by H impurities.[4, 5, 6] This discovery shifted the research interest from intrinsic defects to reactive impurities and their impact on the defect stability and materials properties in general.[4, 7, 8]

Owing to the fully occupied orbitals, it is intuitively thought that noble gases remain inert during the interaction with various materials. Therefore, it is a common belief that the presence of a noble gas (*e.g.*, Ar) during sputtering or wet-chemical synthesis inside a glove box does not affect properties of the product material. Moreover, Ar implantation is often used as a reference to eliminate the influence of implantation damage,[9, 10, 11, 12, 13] what implicitly assumes that Ar defects do not change the materials properties. Despite this, to the best of our knowledge, there is no a single work confirming the negligible role of noble gases in the stability of intrinsic defects and their compensation mechanisms. Keeping in mind the existing hypothesis, we investigate the noble gas defects in wurtzite ZnO with a primary focus on understanding the interaction of He, Ne, and Ar with ZnO:$V_O^q$. The material selection is motivated by a deep knowledge of ZnO:$V_O^q$ and technological importance of ZnO in the semiconductor industry.[14, 15, 16, 17, 18, 19, 20] Despite the focus on ZnO, we also provide common trends that can be extrapolated to other systems.

## RESULTS AND DISCUSSION

**Interaction of noble gases with ideal ZnO:** Utilizing the methodology to screen different insertion sites explained in our previous work,[21] we find that He and Ne in ZnO tend to occupy the octahedral interstitial site shown in Fig. 1a, whereas Ar breaks the Zn-O bonds along the [001] direction (direction *c* in Fig. 1a) and locates between the Zn and O atoms (in-bond site). The gain of the Bader charges on the noble gas atoms is less than −0.07*e*. However, the prediction of partial charges on each atom in a solid is a complex task that is hindered by the overlap of atomic orbitals. In particular, since the Bader charges are calculated by projecting electron density on individual atoms using zero-flux surface approach[22, 23, 24], their interpretation in the analysis of chemical bonding can be misleading. To provide a better understanding of the interaction between the noble gases and ZnO, we calculate spherically averaged charge density rearrangement ($\overline{\Delta\rho}(r)$) around each interstitial noble gas defect as a discrete function of sphere's radius $r$ by averaging the charge density difference ($\Delta\rho$) within the spherical shell of thickness $\Delta r$ = 0.1 Å. Herein, $\Delta\rho = \rho(ZnO:X) - \rho(ZnO) - \rho(X)$, where the charge density ($\rho$) of each component is calculated for the atomic positions of the corresponding ZnO:X (X is He, Ne, or Ar) system. For ZnO:He (Fig. 1b), the charge density near He ($r$ < 0.5 Å) and its nearest surrounding atoms ($r$ > 1.6 Å) increases; while between these two regions it decreases strongly. Similar results are also obtained for the interaction of Ne and Ar with ZnO (Fig. S1). Notably, the electron density is not shifted from noble gas to the surrounding atoms, which shows that He-ZnO interaction in not ionic. Moreover, since the charge density between the noble gas and its surrounding atoms decreases (no electron sharing/donation is observed), the interaction does not have covalent/metallic character either. By analyzing the radial electron distribution for the isolated noble gas atom (Fig. S2) and spherically averaged charge density rearrangement around the defect (Figs. 1b and S1), we conclude that noble gas insertion results in a local destabilization of the electron density for the surrounding atoms (which implies local weakening of Zn-O bonds) caused by minimization of the unfavorable overlap of atomic orbitals of the noble gas and those host atoms. Specifically, the noble gas pushes the electrons of its surrounding atoms away, resulting in the charge density increase near surrounding atoms, while the host material confines the spread of noble gas electron density, leading to charge density increase in the vicinity of the noble gas. Because of this interaction nature, the insertion of noble gases in ZnO costs energy: 1.25, 2.90, and 5.55 eV (7.18 eV for the octahedral interstitial defect) for interstitial He, Ne, and Ar, respectively.

To quantify the overall effect of the noble gas defects on the charge density rearrangements, we calculate a total deviation of the charge density difference for the system as $s = \sum_i^N |\Delta\rho_i| V_i$, where $\Delta\rho_i$ and $V_i$ are charge density difference and volume of the $i$ cell in the discrete grid used for the summation. The computed total deviations of $s$ = 0.28$e$, 0.49$e$, and 0.63$e$ (0.92$e$ for the octahedral interstitial defects) for He, Ne, and Ar imply that the strongest local destabilization of the electron density is caused by Ar insertion, while the smallest effect corresponds to the He defect. Unsurprisingly, larger atomic size of the noble gas results in a stronger repulsion between its atomic orbitals and those of the surrounding atoms. Because of the nature of the noble gas interaction with ZnO, the insertion of larger noble gas results in stronger local destabilization of electron density which correlates with the insertion energies (given above) and atomic displacements. For instance, the shortest He-Zn, Ne-Zn, and Ar-Zn distances are 2.16, 2.23, and 2.24 Å (2.32 Å for the octahedral site), respectively. Indeed, the atomic relaxation is the system's response to the unfavorable overlap of atomic orbitals. The large atomic size and high coordination of the octahedral interstitial site force Ar to break the Zn-O bond and minimize the number of the nearest neighbors and total deviation of the charge density difference. Specifically, Ar at the in-bond site has only 2 neighbors within the distance of 2.60 Å, while it has 12 neighbors at the octahedral site.

The noble gas defects at octahedral sites do not change the electronic structure of ZnO noticeably (Fig. S3). However, Ar in its lowest energy configuration (Fig. 1a) creates an occupied defect state above the valence band maximum (VBM) of the ideal ZnO (Fig. 1c). As an illustration, for 96-atom supercell, the computed energy gap between the defect state and conduction band is 3.12 eV, while the band gap energy of pristine ZnO is 3.42 eV. From the analysis of atom-resolved density of states (DOS) (Fig. 1c) and partial charge density corresponding to the defect state (Fig. 1d), we find that the defect state is occupied by $2e$ and is mainly localized on the undercoordinated O atom (the O atom formed upon Ar insertion breaking Zn-O bond) and its nearest neighbors. These results suggest that the formation of the defect state is due to breaking the Zn-O bond along the [001] direction and charge density redistribution caused by the Ar insertion. Because of this, it can also be speculated that the defect state is a source of photoluminescence that can be used to detect the interstitial Ar defects.

**Interaction of noble gases with ZnO:V$_O$:** Although the noble gas insertion has limited effect on the electronic structure of the pristine ZnO apart from Ar that induces formation of near-VBM defect state, the interaction of the noble gases with vacancy-defected ZnO is different. In fact, the presence of V$_O$ results in a dependence of the insertion energy on distance $d$ between the vacancy and inserted atom (the energy versus distance profile). First, the noble gas can act as an interstitial defect locating at a different distance from V$_O$ (ZnO:V$_O$-X for $d \neq 0$ Å). Second, it can occupy the vacancy site acting as a substitutional defect (ZnO:V$_O$-X for $d = 0$ Å). From the calculations for ZnO:X systems, one can expect that the noble gases tend to occupy the vacancy site to minimize the unfavorable overlap of their atomic orbitals with those of the surrounding atoms: for the substitutional defect, the shortest He-Zn distance is ~2.5 Å, while for the interstitial one it is ~2.2 Å. However, we find that He does not act as a substitutional defect but goes to the interstitial site instead (Fig. 2a). Here, despite a variety of unique interstitial sites, the insertion energy does not depend strongly on the distance to the vacancy. Moreover, for the interstitial sites, the He insertion energy is roughly the same (~1.25 eV) for both the defective and ideal ZnO systems suggesting that the interaction between the noble gas and $V_O$ is short-ranged. To understand these results, we first analyze the electronic structure of ZnO:$V_O$ system and how it is affected by the noble gas insertion (Figs. 2 and S4-S7). As known from the literature[16, 18] and schematically shown in Fig. 2b, the formation of neutral oxygen vacancy leads to the appearance of defect state above the VBM of the host material. This defect state is occupied by 2$e$ and mainly localized in the vicinity of the defect. Therefore, when He occupies the vacancy site, it delocalizes the defect state lifting 2$e$ to the conduction band. According to the analysis of charge density difference (Fig. 2c), akin to X-ZnO interaction, He pushes electrons away from its vicinity, while the host material confines its charge density. Specifically, we observe the increase of charge density near He ($r < 0.2$ Å) and its surrounding atoms ($r > 2.0$ Å). This behavior indicates that the stability of He at the vacancy site is determined by not only the repulsion of atomic orbitals of He and its surrounding atoms but also by the delocalization of otherwise localized vacancy states which costs extra energy; this is in contrast to the stability of the interstitial defect that is determined by the minimization of unfavorable overlap of the atomic orbitals.

The nature of Ne/Ar interaction with ZnO:$V_O$ is similar to that for the He case (Fig. S5-S7). In particular, the formation of the substitutional defects (ZnO:$V_O$-X when $d = 0$ Å) also delocalizes the localized $V_O$ states moving 2$e$ to the conduction band. The interstitial noble gas defects in ZnO:$V_O$ and ideal ZnO systems have roughly the same energetics for all $d \neq 0$ Å. Despite this, the analysis of energy versus distance profiles (Fig. 2a) suggests that, in contrast to He, both Ne and Ar tend to occupy the vacancy site; the lowest insertion energies correspond to $d = 0$ Å. To quantify this effect, for each noble gas, we compute trapping energy as a difference in the insertion energies for interstitial and substitutional defects. The formation of substitutional over interstitial defect becomes more favorable with increase in atomic radius of noble gas; the trapping energies for He, Ne, and Ar are –0.78, 0.74, and 2.71 eV, respectively. Since the energy to delocalize the electrons associated with the Vo is roughly the same for all considered noble gases, the difference in the trapping behavior is caused by the difference in the repulsion of atomic orbitals for the noble gases and their surrounding atoms for the substitutional and interstitial sites. Specifically, for He (the smallest noble gas), the energy gain due to the delocalization of localized Vo states exceeds the energy reduction caused by minimization of the repulsion of atomic orbitals. As a consequence, He acts as an interstitial defect in the ZnO:$V_O$ system. In contrast, for Ne and Ar defects (which at the interstitial configurations destabilize the electron density to a larger extent than He does), the insertion energy can be minimized via the delocalization of localized defect states and formation of the substitutional defects.

**Interaction of noble gases with ZnO:$V_O^q$:** To better understand the interaction between the noble gases and vacancies, we analyze the insertion of noble gases in charged ZnO:$V_O^{1+}$ and ZnO:$V_O^{2+}$ (Figs. 3 and 4) in comparison to that of neutral ZnO:$V_O$ (as shown in Fig. 2). The set of ZnO:$V_O^q$ is considered as it is known from literature[16, 18] that each charged state has own effect on the electronic structures (Fig. S4). For instance, the formation of ZnO:$V_O^{1+}$ creates one occupied and one unoccupied defect states that are located close to the valence and conduction bands of ideal ZnO, respectively (Fig. 3b). Similar to ZnO:$V_O$ (and to pure ZnO), interstitial He and Ne do not change noticeably the electronic structure of the ZnO:$V_O^{1+}$ system, while interstitial Ar produces an in-gap defect state. Moreover, when the noble gas occupies the vacancy side, it delocalizes both defect states, however, lifting up only 1$e$ to the conduction band (Figs. 3b-c). Therefore, for the substitutional noble gas defects, the delocalization of the vacancy states in ZnO:$V_O^{1+}$ system requires less energy than that for the delocalization of 2$e$ in ZnO:$V_O$ system. Hence, all three noble gases (including He) prefer to occupy the charged vacancy site over the formation of the interstitial defects (Fig. 3a).

In contrast to ZnO:$V_O$ and ZnO:$V_O^{1+}$, the formation of ZnO:$V_O^{2+}$ does not create localized defect states (Figs. 4b and S4) as there are no donor electrons associated with the vacancy. Because of this, neither interstitial nor substitutional noble gas defects (except the interstitial Ar defects) alter the electronic structure of ZnO:$V_O^{2+}$ system. In other words, the variation of insertion energy between the interstitial and substitutional defects is mainly determined by the difference in the repulsion of atomic orbitals of the noble gas and its surrounding atoms for the interstitial and substitutional configurations (Figs. 1b and 4c). As a consequence, all three noble gases prefer to occupy the charged vacancy site (Fig. 4a). Moreover, of the considered vacancy charge states, the insertion energies of noble gases in ZnO:$V_O^{2+}$ system are the lowest. For instance, for the substitutional Ar defects in ZnO:$V_O$, ZnO:$V_O^{1+}$, and ZnO:$V_O^{2+}$ systems, the energies are 2.84, 1.39, and 0.95 eV, respectively. Indeed, the comparison of charge density differences for different charged states (Figs. 2c, 3c, 4c and S7) shows that charging the vacancy results in smaller deviation of charge density difference (and hence lower insertion energy) as less electrons are associated with the vacancy site. The distinct insertion energies also suggest that noble gas affects the stability of ZnO:$V_O^q$ systems differently depending on the size of noble gas and number of electrons delocalized by the formation of the substitutional defect. This behavior also noticeably changes the defect transition levels (Fig. S8). It should be noted that although the formation of noble gas defects costs energy, the computed insertion energies for both interstitial and substitutional noble gas defects are comparable to defect formation energies of intrinsic and interstitial H point defects in ZnO.[8, 14, 15, 16, 17, 18, 19] As an illustration, the formation energy of $V_O$ in ZnO under O-rich conditions is 4.41 eV, which is almost 3 times higher than that for interstitial He and only 1.14 eV lower than that for interstitial Ar defect. For the substitutional defects, the insertion energy is as low as 0.13 eV (for ZnO:$V_O^{2+}$-He system). Moreover, despite the energy cost, when the noble gas meets the charged or uncharged vacancy during materials synthesis or noble gas implantation, it can be energetically trapped by the vacancy owing to the large energy difference between the substitutional and interstitial defects. In the other words, to diffuse out from the sample, the noble gas should overcome large energy barrier to move from the vacancy to the interstitial site. As an example, the trapping energies of Ne for ZnO:$V_O$, ZnO:$V_O^{1+}$, and ZnO:$V_O^{2+}$ systems are 0.74, 2.18, and 2.60 eV, respectively.

In summary, our calculations demonstrate that the interaction of the noble gases with ZnO results in the repulsion of atomic orbitals of the noble gases and their surrounding atoms: the noble gas pushes the electron density of its surrounding atoms away from its vicinity, while the host material confines the spread of the noble gas electron density. We thus explain that despite being chemically inert, the noble gas can be a functional dopant. In particular, Ar atom implanted into ideal ZnO breaks the Zn-O bond along [001] direction resulting in the formation of occupied defect states close to the VBM. Our calculations also reveal that the insertion of noble gas can change the stability and electronic structure of ZnO:$V_O^q$. Specifically, when noble gas occupies the vacancy site (*e.g.*, ZnO:$V_O$-Ne for $d$ = 0 Å), it strongly delocalizes the in-gap $V_O^q$ states moving electrons associated with the vacancy to the conduction band. Because of this, the formation of substitutional versus interstitial noble gas defect in ZnO:$V_O^q$ systems is governed by: (I) the size of noble gas in comparison with volume of the vacancy or the interstitial sites determines the energy required for repulsion of neighboring atomic orbitals; (II) the energy cost to delocalize in-gap vacancy states by substitutional noble gas defect. Therefore, He does not tend to occupy the vacancy site in ZnO:$V_O$ system, but instead acts as an interstitial defect. Since the results are generic for materials similar to ZnO, the computed results suggest that despite the fully occupied atomic orbitals the noble gases strongly alter the electronic states of intrinsic defects in technologically important oxides. We thus anticipate that noble gas can serve as a new type of dopants with complementary functionality compared to the traditional dopants in semiconductors.

**METHODS**

All calculations are carried out using the Vienna Ab initio Simulation Package (VASP).[25, 26, 27] To predict the electronic properties with high accuracy, we perform spin-polarized calculations using the originally proposed revised Heyd-Scuseria-Ernzerhof (HSE) screened hybrid functional.[28] Herein, we only modify the fraction of the exact exchange to 0.375 to reproduce the experimental value of ZnO's band gap energy accurately. This methodology is widely accepted for studying defects in ZnO.[16, 19, 20] The search of the defect configurations is performed using the Perdew-Burke-Ernzerhof (PBE) functional.[29] In all calculations, the lattice constants are fixed at the experimental values of $a$ = 3.25 Å and $c$ = 5.21 Å.[30] Atomic relaxations are carried out only at the PBE level until the internal forces are smaller than 0.01 eV/Å. The resulted systems are used directly in the HSE calculations. The defect calculations and the analysis of electronic properties are performed on 96-atom ZnO supercell. The cutoff energies for the plane-wave basis are set to 400 eV for all calculations. 2×2×2 and 4×4×4 Γ-centered Monkhorst-Pack k-grids[31] are used in the Brillouin-zone integrations for the HSE and PBE calculations, respectively. To understand the energetics of the noble gas interaction with the host material, the insertion energy ($E_{ins}$) is calculated as $E_{ins}(ZnO:V_O^q\text{-}X) = \Delta H(ZnO:V_O^q\text{-}X) - \Delta H(ZnO:V_O^q)$, where $\Delta H(ZnO:V_O^q\text{-}X)$ and $\Delta H(ZnO:V_O^q)$ are the defect formation energies for the system containing both the O vacancy as well as the noble gas X and the system with O vacancy only, respectively. As it is easy to see, in the dilute limit, $E_{ins}(ZnO:V_O^q\text{-}X) = E(ZnO:V_O^q\text{-}X) - E(ZnO:V_O^q) - E(X)$, where $E(ZnO:V_O^q\text{-}X)$, $E(ZnO:V_O^q)$, and $E(X)$ are energies of the corresponding system. $E(X)$ is calculated for the isolated noble gas X. Correspondingly, the insertion energy of the noble gas in ideal ZnO is taken to be equal to the formation energy of the interstitial noble gas defect, $E_{ins}(ZnO:X) = E(ZnO:X) - E(ZnO) - E(X)$. The defect formation energies and finite size corrections are calculated within the methodology described by Lany and Zunger[32, 33] as implemented in the pylada-defects code.[34] To calculate the insertion energy versus distance profile, we take oxygen vacancy position as the position of noble gas in the substitutional configuration, which can result in up to 0.1 Å inaccuracy for the distance between the vacancy and noble gas defect. The trapping energy ($E_{trap}$) is calculated as $E_{trap} = E_{ins}(ZnO:X) - E_{ins}(ZnO:V_O^q\text{-}X$ for $d$ = 0 Å). It should be noted we use the term trapping energy (not binding or clustering energy) since the change in charge density does not lower the total energy, which however is the case for a traditional defect where chemical bonding is responsible for the interaction. The results for the energetics and electronic properties (*i.e.,* DOS) are presented for the HSE calculations. The analysis of Bader charges, charge density difference, spherically averaged charge density rearrangements, and total deviation of the charge density difference are performed at the PBE level. The Bader charges are calculated within the scheme developed by Henkelman *et al.*[22, 23, 24] and presented with respect to that for the neutral isolated atoms. The obtained results are analyzed using Vesta[35] and pymatgen.[36]

## DATA AVAILABILITY

All data needed to evaluate the conclusions in the paper are present in the paper and the Supplementary Materials. Additional data for this study are available from the corresponding author upon request.

## ACKNOWLEDGMENTS

This work is financially supported by the Research Council of Norway (ToppForsk project: 251131). We acknowledge Norwegian Metacenter for Computational Science (NOTUR) for providing access to supercomputer resources.

## AUTHOR CONTRIBUTIONS

O.I.M and C.P designed the project. O.I.M performed calculations and wrote most of the paper. O.I.M and K.S analyzed results. All co-authors participated in the discussions and manuscript preparation. C.P supervised the theoretical studies.

## ADDITIONAL INFORMATION

Supplementary information accompanying the paper is available.

**COMPETING INTERESTS:** the authors declare no competing interests.

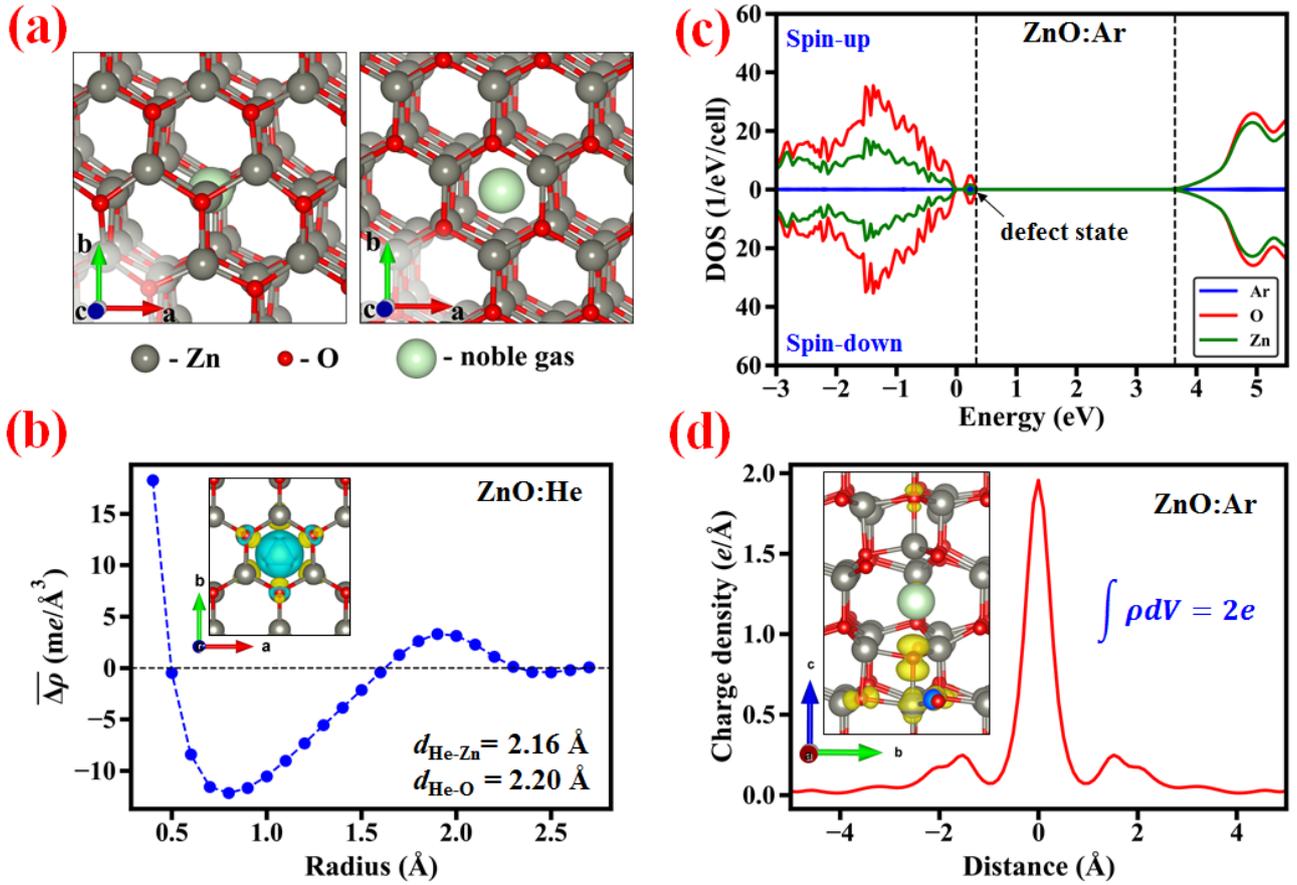

**Figure 1**. Insertion of noble gases in pure ZnO. (a) Schematic illustration of the lowest energy sites for Ar (left) and He/Ne (right). (b) Spherically averaged charge density rearrangement ($\overline{\Delta\rho}(r)$) around He computed by averaging charge density difference within the spherical shell of thickness $\Delta r$ = 0.1 Å. The inset shows the charge density difference for ZnO:He system. The charge density difference is defined as $\Delta\rho = \rho(ZnO:He) - \rho(ZnO) - \rho(He)$, where the charge density ($\rho$) of each component is calculated for the atomic positions of ZnO:He system. The blue and yellow regions represent the charge density reduction and charge density increase regions, respectively (the isosurface is set to 6.75 m$e$/Å$^3$). (c) DOS for the lowest energy configuration of ZnO:Ar; the DOS for the conduction band is multiplied by 40 for better visualization. The vertical dashed lines mark the highest occupied and lowest unoccupied states, independent of the spin component. (d) Partial charge density (yellow regions) corresponding to the defect state induced by Ar insertion (the isosurface is set at 67.5 m$e$/Å$^3$) and its projection on the axis $a$.

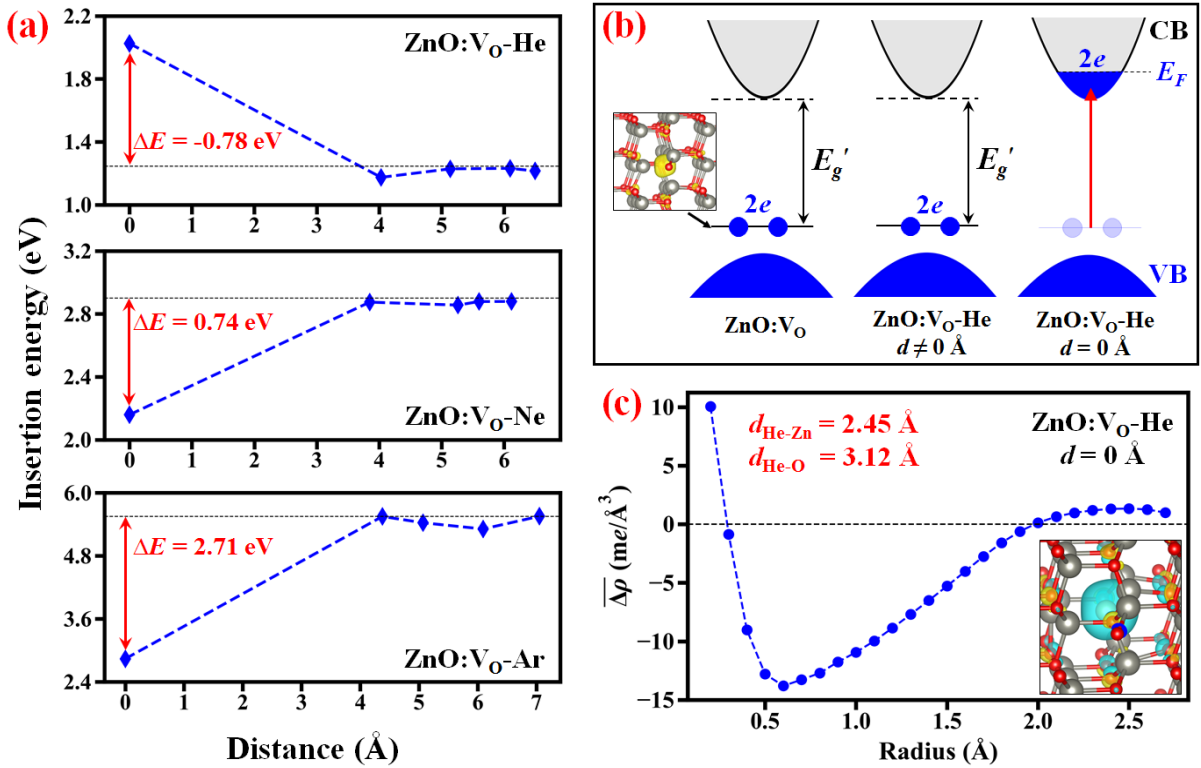

**Figure 2**. Interaction of noble gases with ZnO:$V_O$. (a) Insertion energy as a function of the distance between the vacancy and the noble gas. For each system, the trapping energy (Δ$E$) is calculated as the difference of insertion energies for the interstitial and substitutional defects. (b) Schematic illustration showing how He insertion changes the electronic states of ZnO:$V_O$ system for the case of interstitial and substitutional defect. The inset depicts partial charge density at the defect state (the isosurface is set to 67.5 m$e$/Å$^3$). The up down arrow marks the energy gap ($E_g'$) between the highest occupied and lowest unoccupied states. (c) Spherically averaged charge density rearrangement ($\overline{\Delta\rho}(r)$) around the substitutional He defect computed by averaging charge density difference within the spherical shell of thickness Δ$r$ = 0.1 Å. The inset shows the charge density difference for ZnO:$V_O$-He ($d$ = 0 Å) computed as Δ$\rho$ = $\rho$(ZnO:$V_O$-He) − $\rho$(ZnO:$V_O$) − $\rho$(He), where the charge density ($\rho$) of each component is calculated for the atomic positions of ZnO:He system. The blue and yellow regions represent the charge density reduction and charge density increase regions, respectively. The isosurface is set at 6.75 m$e$/Å$^3$.

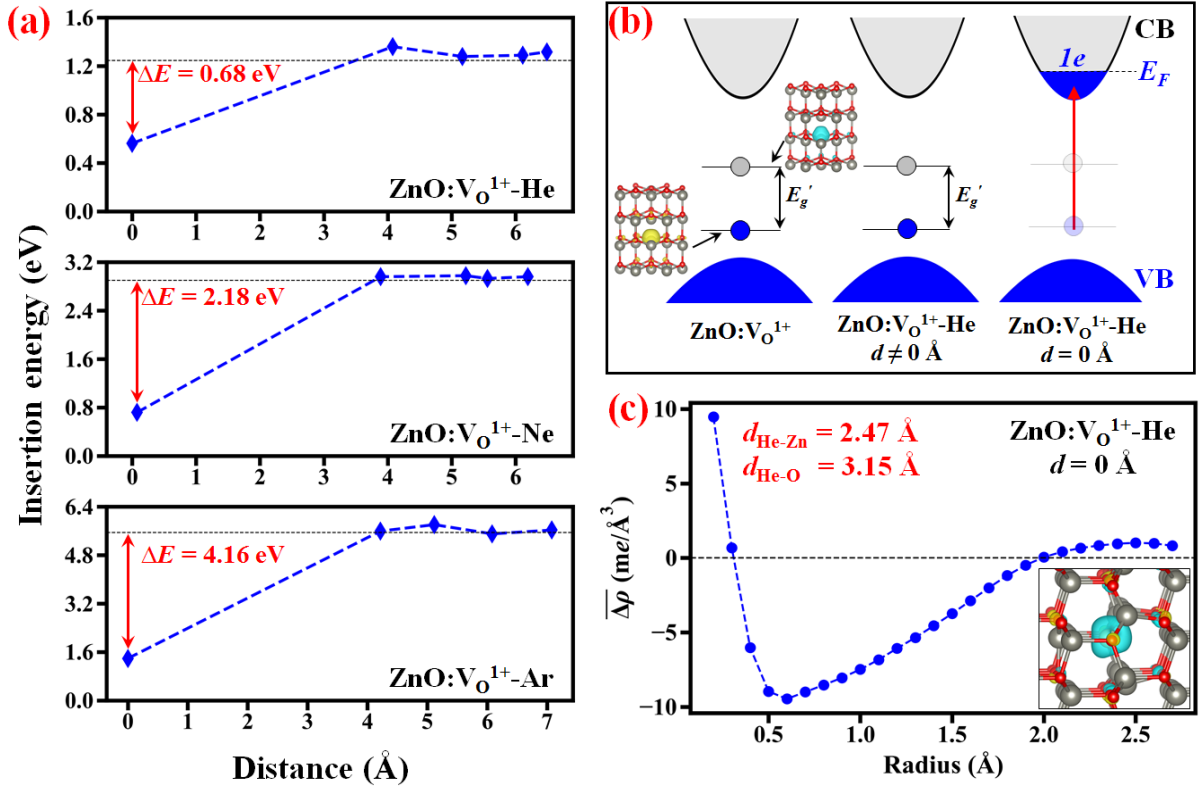

**Figure 3.** Interaction of noble gases with ZnO:$V_O^{1+}$. (a) Insertion energy as a function of the distance between the vacancy and the noble gas. For each system, the trapping energy (ΔE) is calculated as the difference of insertion energies for the interstitial and substitutional defects. (b) Schematic illustration showing how He insertion changes the electronic states of ZnO:$V_O^{1+}$ system for the case of interstitial and substitutional defect. The inset depicts partial charge density at the defect states (the isosurface is set to 33.75 m$e$/Å$^3$). The up down arrow marks the energy gap ($E_g^{'}$) between the highest occupied and lowest unoccupied states. (c) Spherically averaged charge density rearrangement ($\overline{\Delta\rho}(r)$) around the substitutional He defect computed by averaging charge density difference within the spherical shell of thickness Δ$r$ = 0.1 Å. The inset shows the charge density difference for ZnO:$V_O^{1+}$-He ($d$ = 0 Å) computed as Δ$\rho$ = $\rho$(ZnO:$V_O^{1+}$-He) – $\rho$(ZnO:$V_O^{1+}$) – $\rho$(He), where the charge density ($\rho$) of each component is calculated for the atomic positions of ZnO:$V_O^{1+}$-He system. The blue and yellow regions represent the charge density reduction and charge density increase regions, respectively. The isosurface is set to 6.75 m$e$/Å$^3$.

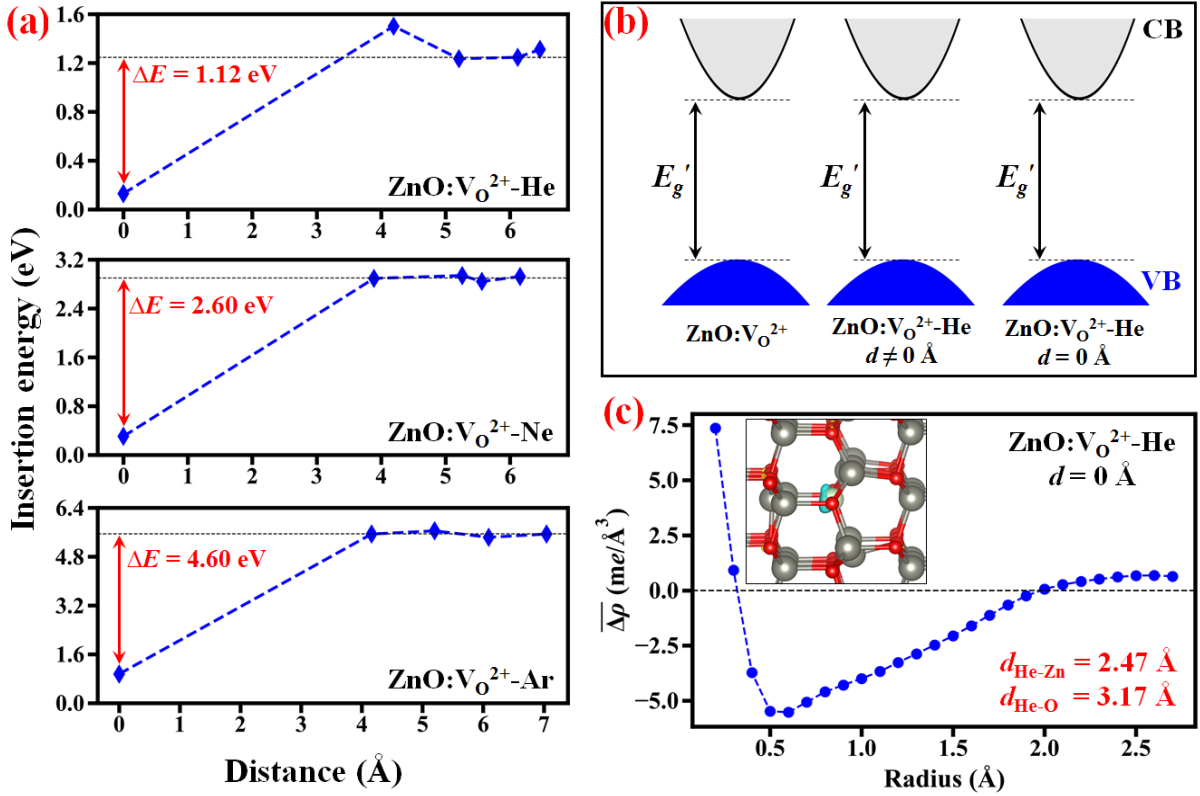

**Figure 4**. Interaction of noble gases with ZnO:$V_O^{2+}$. (a) Insertion energy as a function of the distance between the vacancy and the noble gas. For each system, the trapping energy (ΔE) is calculated as the difference of insertion energies for the interstitial and substitutional defects. (b) Schematic illustration showing how He insertion changes the electronic states of ZnO:$V_O^{2+}$ system for the case of interstitial and substitutional defect. The up down arrow marks the energy gap ($E_g'$) between the highest occupied and lowest unoccupied states. (c) Spherically averaged charge density rearrangement ($\overline{\Delta\rho}(r)$) around the substitutional He defect computed by averaging charge density difference within the spherical shell of thickness Δr = 0.1 Å. The inset shows the charge density difference for ZnO:$V_O^{2+}$-He (d = 0 Å) computed as Δρ = ρ(ZnO:$V_O^{2+}$-He) − ρ(ZnO:$V_O^{2+}$) − ρ(He), where the charge density (ρ) of each component is calculated for the atomic positions of ZnO:$V_O^{2+}$-He system. The blue and yellow regions represent the charge density reduction and charge density increase regions, respectively. The isosurface is set to 6.75 m$e$/Å$^3$.

15